\documentclass[reprint,prd,twocolumn,preprintnumbers,superscriptaddress,dblfloatfix]{revtex4-1}
\usepackage{graphicx}   
\usepackage{dcolumn}    
\usepackage{bm}         
\usepackage{amssymb}    
\usepackage{setspace}
\usepackage{amsmath, amssymb, setspace}
\usepackage{array}
\usepackage{booktabs}
\usepackage{mathrsfs}
\usepackage{indentfirst}
\usepackage{slashed}
\usepackage{float}
\usepackage{lmodern}
\usepackage{hyperref}
\usepackage{xcolor}
\usepackage{multirow}
\usepackage{epstopdf}
\usepackage{ulem}
\usepackage{tabularx}
\usepackage[justification=raggedright]{caption}
\usepackage[flushleft]{threeparttable}
\usepackage{hyperref}
\usepackage{soul}
\usepackage[utf8]{inputenc}

\begin{document}

\preprint{IPMU18-0160}

\title{Primordial Black Holes and the String Swampland}

\author{Masahiro Kawasaki}
\email[]{kawasaki@icrr.u-tokyo.ac.jp}
\affiliation{
Institute of Cosmic Ray Research, University of Tokyo, Kashiwa, 277-8582, Japan}
\affiliation{Kavli IPMU (WPI), UTIAS, University of Tokyo, Kashiwa, 277-8583, Japan}

\author{Volodymyr Takhistov}
\email[]{vtakhist@physics.ucla.edu}
\affiliation{Department of Physics and Astronomy, University of California, Los Angeles\\
Los Angeles, California, 90095-1547, USA}

\date{\today}

\begin{abstract}
The ``swampland conjectures'' have been recently suggested as a set of criteria to assess if effective field theories (EFTs) are consistent with a quantum gravity embedding. Such criteria, which restrict the behavior of scalar fields in the theory, have strong implications for cosmology in the early universe. As we demonstrate, they will also have direct consequences for formation of primordial black holes (PBHs) and dark matter (DM). 
\end{abstract}

\maketitle

\section{Introduction} 

Primordial black holes can form in the early universe and can account for all or part of the dark matter~(e.g.~\cite{Zeldovich:1967,Hawking:1971ei,Carr:1974nx,GarciaBellido:1996qt,Khlopov:2008qy,Frampton:2010sw,Kawasaki:2016pql,Cotner:2016cvr,Carr:2016drx,Inomata:2016rbd,Inomata:2017okj,Garcia-Bellido:2017aan,Inoue:2017csr,Georg:2017mqk,Inomata:2017bwi,Kocsis:2017yty,Ando:2017veq}). They have also been linked to a variety of topics in astronomy, including the recently discovered \cite{Abbott:2016blz,Abbott:2016nmj,Abbott:2017vtc} gravitational waves \cite{Nakamura:1997sm,Clesse:2015wea,Bird:2016dcv,Raidal:2017mfl,Eroshenko:2016hmn,Sasaki:2016jop,Clesse:2016ajp,Takhistov:2017bpt}, formation of supermassive black holes~\cite{Bean:2002kx,Kawasaki:2012kn,Clesse:2015wea}, cosmic infrared background fluctuations \cite{Kashlinsky:2016sdv} as well as  $r$-process nucleosynthesis~\cite{Fuller:2017uyd}, gamma-ray bursts and  micro-quasars \cite{Takhistov:2017nmt} from compact star disruptions.

The vast ``landscape'' of string theory vacua
is believed to result in EFTs consistent with quantum gravity. On the other hand, the ``swampland'' contains EFTs for which this is not the case \cite{Vafa:2005ui}.
Recently, two conditions have been proposed, the so-called ``swampland conjectures'', to discriminate between these two classes:
\begin{itemize}
\item \textit{SC1 \cite{Ooguri:2006in}:} scalar field excursion, measured in Planck units in the field space, is bounded from above
\begin{equation}
 |\Delta \phi| \lesssim  d \sim \mathcal{O}(1)
\end{equation}
\item \textit{SC2 \cite{Obied:2018sgi}:} the gradient of the potential of
a canonically normalized scalar field satisfies
\begin{equation}
 \dfrac{|V^{\prime}|}{V} \gtrsim c \sim \mathcal{O}(1)
\end{equation}
\end{itemize}
Here, $c,d$ are constants of order unity.
We take the Planck mass to be $M_{pl} (\equiv 2.4 \times 10^{18}~\text{GeV})  = 1$ throughout.
As discussed in \cite{Agrawal:2018own}, the above criteria have profound implications for the early universe cosmology, as follows. The general features of inflationary physics can be parametrized by the slow-roll parameters $\epsilon$ and $\eta$, which in terms of the scalar inflaton potential are given by~\cite{Liddle:1994dx}
\begin{equation}
\epsilon = \dfrac{1}{2} \Big(\dfrac{V^{\prime}}{V}\Big)^2~~~,~~~\eta = \Big(\dfrac{V^{\prime\prime}}{V}\Big)~.
\end{equation}
For a successful period of inflation one requires $\epsilon, |\eta| \ll 1$. The first slow-roll parameter $\epsilon$
is related to the elapsed number of expansion e-folds $N$, with $d N = H dt$ and $H$ denoting the Hubble parameter, as $|d \phi/d N| = \sqrt{2  \epsilon}$. Taking that inflation has lasted at least 60 e-folds to address the problems with the Big Bang cosmology, one obtains $60 < d/c $, which is in mild tension with $d, c \sim \mathcal{O}(1)$.
The tensor-to-scalar ratio $r = 16 \epsilon$, constrained by the cosmic microwave background (CMB) $B$-modes as $r < 0.07$ at the comoving wave-number pivot scale of $k_0 = 0.05$ Mpc$^{-1}$ by the \textit{Planck}-2018 satellite data \cite{Aghanim:2018eyx}, leads to $\Delta \phi \lesssim 6$, which approaches the bound implied by \textit{SC1}.
The precise values of $c, d$ depend on the details of string compactification and can deviate from strict unity \cite{Dias:2018ngv}. As has also been noted in \cite{Agrawal:2018own} and further explored in e.g. \cite{Heisenberg:2018yae,Denef:2018etk}, the swampland conjectures will also have strong implications for dark energy.

PBHs form  when density fluctuations are comparable to $\mathcal{O}(1)$ at the horizon
crossing. Hence, for PBHs to constitute dark matter,  one
requires a large amplification of the inflationary power
spectrum between the cosmic microwave background (CMB)
 and the PBH mass scales. As we demonstrate, the swampland criteria have direct consequences for formation of PBHs and dark matter.

\section{Primordial Black Hole Formation}

The power spectrum of primordial curvature perturbations  is given by  (e.g. \cite{Kosowsky:1995aa})
\begin{equation}
\Delta_{\zeta}^2(k) = \dfrac{k^3 P_{\zeta}(k)}{2 \pi^2} = A_s \Big(\dfrac{k}{k_0}\Big)^{n_s - 1}~,
\end{equation}
where $A_s = (2.105 \pm 0.030
)  \times 10^{-9}$ is the scalar power spectrum amplitude and $n_s = 0.9665 \pm 0.0038 $ is the scalar spectral index, evaluated from the \textit{Planck}-2018 measurements at $k_0$ \cite{Aghanim:2018eyx}. 
The PBH mass is defined to be $M = \gamma M_H$, where $\gamma$ is an $\mathcal{O}(1)$ parameter specifying efficiency of overdensity collapse to a black hole and $M_H = 1/2 G H$ is the horizon mass. The corresponding scale $k_M = a_H H = a_{exit} H_{\rm inf}$ has exited $N$ e-folds after the CMB scale $k_0 = a_0 H_{\rm inf}$, where $H_{\rm inf}$ is the Hubble parameter value during inflation. Taking $H_{\rm inf} \simeq const$, one obtains~\cite{Motohashi:2017kbs}
\begin{equation} \label{eq:nhor}
N = 18.4 - \dfrac{1}{12} \log \Big(\dfrac{g_{\ast}}{g_{\ast 0}}\Big) + \dfrac{1}{2} \log \gamma - \dfrac{1}{2} \log \Big(\dfrac{M}{M_{\odot}}\Big)~,
\end{equation}
where $g_{\ast}$ denotes the effective degrees of freedom in the energy density, with $g_{\ast 0} = 3.36$ being their number today.

For PBHs to constitute dark matter, the minimal mass that is necessary in order to survive Hawking evaporation to the present day is given by 
\begin{equation}
M_{\rm min} = 1.5 \times 10^{-21} \Big(\dfrac{\Omega_m h^2}{0.14}\Big)^{-2/3} M_{\odot}~,
\end{equation}
where $\Omega_m h^2 = 0.14240 \pm 0.00087$ from \textit{Planck}-2018 \cite{Aghanim:2018eyx}. Hence, this scale left the horizon $N \simeq 42$ e-folds after the CMB, where in Eq.~\eqref{eq:nhor} the values of $g_{\ast} = 106.75$ as in the Standard Model and  $\gamma = 1$ have been assumed, conservatively.

Starting from the usual Press-Schechter formalism \cite{Press:1973iz} for PBH formation during the  radiation-dominated era, for all of the DM to reside in PBHs of mass $M > M_{\rm min}$ one needs \cite{Motohashi:2017kbs}
\begin{equation} \label{eq:curvpbh}
\Delta_{\zeta}^2 (M_{\rm min}) \simeq 2.1 \times 10^{-2}~.
\end{equation}
 Modification of $\Delta_{\zeta}^2 (M_{\rm min})$ by an order of magnitude, as suggested by the recent analysis of \cite{Germani:2018jgr}, will not have a drastic effect on our conclusions.

At the leading order in the slow-roll, the curvature and tensor perturbations, respectively, are given by
\begin{equation} \label{eq:curvp}
\Delta_{\zeta}^2 \simeq \dfrac{H_{\rm inf}^2}{8 \pi^2 \epsilon}~~~,~~~\Delta_t^2 \simeq \dfrac{2 H_{\rm inf}^2}{\pi^2}~.
\end{equation}
Using the observed value of $\Delta_{\zeta}^2(k_0) \simeq 2.1 \times 10^{-9} $ \cite{Akrami:2018odb}, the tensor-to-scalar ratio can be parametrized as 
\begin{equation}
r = \dfrac{\Delta_t^2}{\Delta_{\zeta}^2} \simeq  ~ 9.6 \times 10^{7} H_{\rm inf}^2~.  
\end{equation}
Eliminating $H_{\rm inf}$ from Eq.~\eqref{eq:curvp} and substituting the required perturbation amplification for PBHs, as given by Eq.~\eqref{eq:curvpbh}, we obtain
\begin{equation}
\epsilon = 6.3 \times 10^{-9} r~.
\end{equation}
Hence, taken together with the constraint from \textit{Planck}-2018 of $r < 0.07$ \cite{Akrami:2018odb}, PBH formation consistent with the CMB measurements restricts the first slow-roll parameter $\epsilon$ to be
\begin{equation} \label{eq:pbh1sr}
\epsilon < 4.4 \times 10^{-10}~.
\end{equation}

The required amplification for PBHs to constitute DM also leads
to $\mathcal{O}(1)$ violation of the slow-roll parameter combination, irrespective of the inflationary model details. Namely, given the required amplification of curvature perturbations to form PBHs for DM over $N = 42$ e-folds after the CMB, using Eq.~\eqref{eq:curvp}, one obtains \cite{Motohashi:2017kbs}
\begin{equation} \label{eq:pbhviol}
\Big| \dfrac{\Delta \log \epsilon}{\Delta N}\Big| > 0.4~.
\end{equation}
Since the horizon-flow equations \cite{Kinney:2002qn,Hoffman:2000ue} give
\begin{equation} \label{eq:flow}
\dfrac{d \log \epsilon}{d N}   =  2 \Big[\Big(\dfrac{V^{\prime}}{V}\Big)^2 -  \dfrac{V^{\prime\prime}}{V}\Big] =  4 \epsilon - 2\eta~,
\end{equation}
from Eq.~\eqref{eq:pbhviol} we have
\begin{equation} \label{eq:pbh2sr}
|2 \epsilon - \eta| > 0.2~.
\end{equation}
Together with Eq.~\eqref{eq:pbh1sr}, this can viewed as a restriction on the second slow-roll parameter $\eta$ for PBH DM, consistent with CMB observations.

As discussed, significant number of long-lived PBHs require power enhancement on smaller scales, corresponding to large wave-number $k$. This demands that the spectral index is running and is ``blue-tilted'', with $n_s > 1$ at relevant scales (e.g. \cite{Josan:2010cj,Drees:2011yz}). In terms of the slow-roll parameters, this translates to 
\begin{equation} \label{eq:specindex}
n_s - 1 = 2 \eta - 6 \epsilon > 0~.
\end{equation}

Here we comment on the validity of Eq.~(\ref{eq:curvp}).
In deriving Eq.~(\ref{eq:curvp}) we have used the slow-roll approximation, which may not be applicable, as suggested by Eq.~(\ref{eq:pbhviol}).
In fact, \cite{Motohashi:2017kbs} shows that naive use of Eq.~(\ref{eq:curvp}) leads to some errors in PBH formation models.
However, the errors are not so large as to affect our argument.  

We note, in passing, that PBHs can also form in matter-dominated era (e.g. \cite{Harada:2016mhb}), which requires that the collapsing regions are sufficiently spherically symmetric.

\section{Swampland Restriction}

From the first slow-roll parameter $\epsilon$, combining $\epsilon\gtrsim c^2/2$ from \textit{SC2} and Eq.~\eqref{eq:pbh1sr} for PBH formation, one obtains
\begin{equation} \label{eq:pbhsw1}
c \lesssim 3.0 \times 10^{-5}~.
\end{equation}

The swampland conjectures will also  constrain the second slow-roll parameter $\eta$. 
Since PBH formation implies that the spectrum is blue-tilted at the relevant scales, 
we restrict ourselves to $V'' > 0$ potential, resulting in $\eta > 0$.
Then, \textit{SC2} leads to~\cite{Kinney:2018nny}
\begin{equation} \label{eq:scetares}
\eta \gtrsim c^2.
\end{equation}
From the blue-tilted spectrum requirement of Eq.~\eqref{eq:specindex}, combined with \textit{SC1}, one obtains a slightly stronger restriction of 
\begin{equation}
\eta > 3\epsilon \gtrsim \dfrac{3 }{2} c^2~,
\end{equation}
consistent with Eq.~\eqref{eq:scetares}.
Hence, 
\begin{equation}\label{eq:pbhsw2}
\Big|\dfrac{d \log \epsilon}{d N}\Big| = |4\epsilon - 2 \eta| > 2\epsilon \gtrsim c^2~,
\end{equation}
resulting in $|2 \epsilon - \eta| > c^2/2$, which is to be compared with~Eq.~\eqref{eq:pbh2sr}.

While the swampland criteria with $c \sim \mathcal{O}(1)$ automatically satisfies the restriction on the second parameter $\eta$, it is strictly incompatible with the range of the first slow-roll parameter $\epsilon$ as required for PBH DM.  

We note that it is possible to ease the restrictions of the swampland criteria, for example,  by considering a multi-field inflationary setup \cite{Achucarro:2018vey,Garg:2018reu}, curvaton models \cite{Kehagias:2018uem}, models with non-canonical kinetic terms \cite{Das:2018hqy} (e.g. $k$-inflation \cite{ArmendarizPicon:1999rj}) or that fluctuations begin in an excited initial state and not the Bunch-Davies vacuum \cite{Brahma:2018hrd}. Here, the relationship between the slow-roll parameters in  curvature perturbations as well as other quantities will be modified. Discussion of PBH formation in that context will be treated elsewhere. The tension with the swampland conjectures could be also weakened by modifying the proposed criteria themselves, see \cite{Andriot:2018wzk,Garg:2018reu} for potential suggestions.

\section{Conclusions}

We have shown that the swampland conjectures, as originally proposed, are incompatible with formation of PBHs that can constitute DM in the context of single-field inflation.~This highlights that
placing restrictions on the behavior of the scalar fields in EFTs can have significant implications for structure formation in the early universe as well as dark matter. 

\section*{Acknowledgements}

The work of M.K. was supported by Japan Society for
the Promotion of Science KAKENHI Grants
No. 17H01131 and No. 17K05434, Ministry of
Education, Culture, Sports, Science and Technology of Japan (MEXT) KAKENHI Grant No. 15H05889, and
World Premier International Research Center Initiative (WPI Initiative), MEXT, Japan. The work of V.T. was supported by the U.S. Department of Energy Grant
No. DE-SC0009937.

\bibliography{pbhsw}

\end{document}